\documentclass[twocolumn,showpacs,aps,pra]{revtex4}

\usepackage{graphicx}
\usepackage{amsfonts, amsmath, amstext, amssymb, amsfonts, amsxtra}
\usepackage{braket}
\usepackage{bbm}
\usepackage{bbold}
\usepackage[protrusion=true,expansion=true]{microtype}
\usepackage{wasysym}
\usepackage{cleveref}
\usepackage{color}
\usepackage{ulem}

\newcommand{\im}{{i}}         
\newcommand{\id}{\mathbb{1}}  

\begin{document}

\title{Control of a single-particle localization in open quantum systems}
\author{O.S.~Vershinina$^{1}$,  I.I.~Yusipov $^{2}$, S.~Denisov $^{2,3}$, M.V.~Ivanchenko $^{2}$, and T.V.~Laptyeva $^{1,2}$}

\affiliation{ $^{1}$ Institute of Supercomputing Technologies, Lobachevsky State University of Nizhny Novgorod, Russia \\
$^{2}$ Institute of Information Technologies, Mathematics and Mechanics, Lobachevsky State University of Nizhny Novgorod, Russia \\
$^{3}$ Department of Theoretical Physics, University of Augsburg, Germany}


\begin{abstract}
We investigate the possibility to control localization properties of the 
asymptotic state of an open quantum system with a tunable synthetic dissipation. 
The  control  mechanism relies on the matching between properties of dissipative  operators, acting on neighboring sites and specified by a single control parameter,
and the spatial phase structure of eigenstates  of the system Hamiltonian. As a result, the latter coincide (or near coincide) with the dark states of the 
operators. In a disorder-free Hamiltonian with a flat band, one can either obtain a localized  asymptotic state or populate whole
flat and/or dispersive bands, depending on the value of the control parameter. 
In a disordered Anderson system,  the asymptotic state 
can be localized anywhere in the spectrum of the Hamiltonian. The dissipative control is robust with respect to an additional local dephasing.
\end{abstract}

\maketitle

%


{\it Introduction. --} Anderson localization \cite{anderson} is a hallmark of modern physics \cite{Kramer1993,Evers2008,fifty}, which 
has been observed in experiments with light, sound and matter waves \cite{fifty2,Segev2013,Billy2008,Roati2008,Kondov2011,Jen2012}. 
However, the research activity in the field  of \textit{open} (i.e., interacting with their environments)
quantum systems \cite{book}, has, to some extent, bypassed this phenomenon. 
Since Anderson localization is based on long-range interference effects, it is intuitive to expect that
dissipation will blur the latter and thus eventually  destroy the former. Another example of localization, compact 
states induced by the flat band topology \cite{Bergman2008,Richter2006,Flach2014}, also involves destructive 
interference and therefore might be expected to be fragile with respect to any kind of  dissipation.  

During the last decade, it was realized that dissipative effects are not always a nuisance, especially when they can be controlled. The idea that 
a synthetic  dissipation can be used to bring many-body systems into pure and highly entangled states
\cite{DiehlZoller2008, KrausZoller, wolf2009}, lead to the creation of a field conventionally called `dissipative engineering'.  
However, in the context of localization, it remained unknown until very recently  whether signatures of localization can 
survive in an open quantum system at the asymptotic limit. Recent  studies  of Anderson 
localization in the presence of dephasing effects confirmed its destruction in the asymptotic limit, 
although revealed that its footprints can still  be detected on the way to the a completely de-localized asymptotic state \cite{les0}. 

There were some clues that localization 
can survive in semi-classical and classical systems \cite{Stano2013, LiuJ.2014,ivanchenko2015a,ivanchenko2015b}, which motivated us to 
search for the possibility to detect signatures of Anderson localization in open quantum systems. In Ref. 
~ \cite{Yusipov2017}  it was demonstrated that  pairwise, i.e., acting on a pair of neighboring  sites only, dissipative operators (dissipators) can favor
Anderson modes from a particular designated part of the spectrum. 
Depending on the value of a control phase parameter, dissipators sculpture the asymptotic state 
by selecting  modes either from the lower or upper band edge of the spectrum or from the spectrum center.

In this paper we investigate the possibility to control localization properties of the asymptotic states in single-particle systems,
where localization in the Hamiltonian limit is induced by flat band topology or by disorder. We demonstrate that localization signatures 
survive even in the presence of the most severe type of dissipation, that is a local dephasing. Another word, by introducing  a synthetic dissipation
into  already open (to an uncontrollable locally acting decoherence) system with Anderson of a flat-band Hamiltonian, one can observe footprints of localization
in the system asymptotic state.


{\it Tunable dissipative operators. --} We address the evolution of an open $N$-dimensional quantum system whose density operator $\varrho$ governed by the Lindblad  master equation 
\cite{book,alicki},
\begin{equation}
\label{eq:1}
\dot{\varrho} = \mathcal{L}(\varrho) = -\im [H,\varrho] + \mathcal{D}(\varrho).
\end{equation}
The first term on the r.h.s.\ captures the unitary evolution of the system
governed by tight-binding lattice Hamiltonian
\begin{align}
H=&\sum_j \epsilon_j b_j^{\dagger}b_j -\sum\limits_{m\in\mathcal{N}(j)}b_{j}^{\dagger} b_{m}, 
\label{eq:2}
\end{align}
where $\epsilon_j$ are on-site energies, $\mathcal{N}(j)$ is the set of neighbors of the $j$-th site, determined by the lattice topology, $b_j$ and $b_j^{\dagger}$ 
are the annihilation and creation operators of a boson on the $j$-th site. Periodic boundary conditions are imposed.

The  dissipative part of the Lindblad generator $\mathcal{L}$,
\begin{equation}
\label{dissipator}
\mathcal{D}(\varrho) = \sum_{j=1}^{S} \gamma_{j}(t) \left[V_j\varrho V^\dagger_j - \frac{1}{2}\{V^\dagger_jV_j,\varrho\}\right]
\end{equation}
is built from the set of $S$ operators, $\{V_j\}_{1,...,S}$, which capture action of the environment on the system.

The asymptotic density operator $\varrho_{\infty}$ is the product of  the joint action of the Hamiltonian and dissipative operators. 
If all  operators are Hermitian, $V_j \equiv V^\dagger_j$, the asymptotic state
is universal and trivial; namely, the density operator is the normalized identity $\varrho_{\infty}  = \id/N$, irrespective of properties of the Hamiltonian. 
An example is the on-site dephasing operator, 
\begin{equation}
V_j^{d}=b_{j}^{\dagger}b_{j},
\label{eq:4a}
\end{equation}
one of the  most popular choices to study relaxation processes in open systems, single \cite{les0} and many-particle ones \cite{fish,les,les2,lazz}. 



We also consider non-Hermitian dissipative operators that act on pairs of neighboring sites,
\begin{equation}
V_{j,n(j)}^{nn}=(b_{j}^{\dagger} + e^{i \alpha}b_{n(j)}^{\dagger})(b_{j}-e^{-i\alpha}b_{n(j)}),
\label{eq:4}
\end{equation}
where $n(j) \in \mathcal{N}(j)$ is the index of a neighbor of $j$-th site.
This  type of dissipation, for a one-dimensional chain, $n(j)=j+1$,   and in the  many-body context, was introduced in Refs.~\cite{DiehlZoller2008,KrausZoller}, where
it was shown to be able to bring the system into the BEC state. In Ref. \cite{Yusipov2017} these dissipators were used in the context of Anderson model,
to create non-trivial asymptotic states featuring signatures of localization. 
Operators (5) are parametrized by a phase $\alpha$, which makes the dissipation  phase-selective. 
For example, when $\alpha=0$, the corresponding  operators try  to synchronize the dynamics on the neigboring sites, 
by constantly recycling anti-symmetric out-of-phase mode into the symmetric in-phase one; the effect of $\alpha=\pi$ is the opposite.

A physical implementation of a Bose-Hubbard chain with neighboring sites coupled by such  dissipators
was discussed in Ref.~ \cite{marcos2012}. The proposed set-up consists of an array of superconductive resonators 
coupled by qubits; a pair-wise dissipator with $n(j)=j+1$ and arbitrary phase $\alpha$ can be realized with this set-up by adjusting  position of the qubits
with respect to the centers of  the corresponding cavities. 
More specifically, the relative position of a qubit in a cavity controls the phase of a complex coupling constant $q_j$ in the Jaynes-Cummings coupling term, 
$q_j^*b_j^{\dagger}\sigma_j^- + q_jb_j \sigma_j^+$, where a qubit operator $\sigma_j^- = \arrowvert g_j\rangle\langle e_j|$ \cite{QO}. 
To implement dissipator  with $\alpha \neq 0$,  the coupling constant should vary  as $q_j = |q|\exp(-i\alpha j)$.

\begin{figure}[t]
(a) \includegraphics[width=0.9\columnwidth]{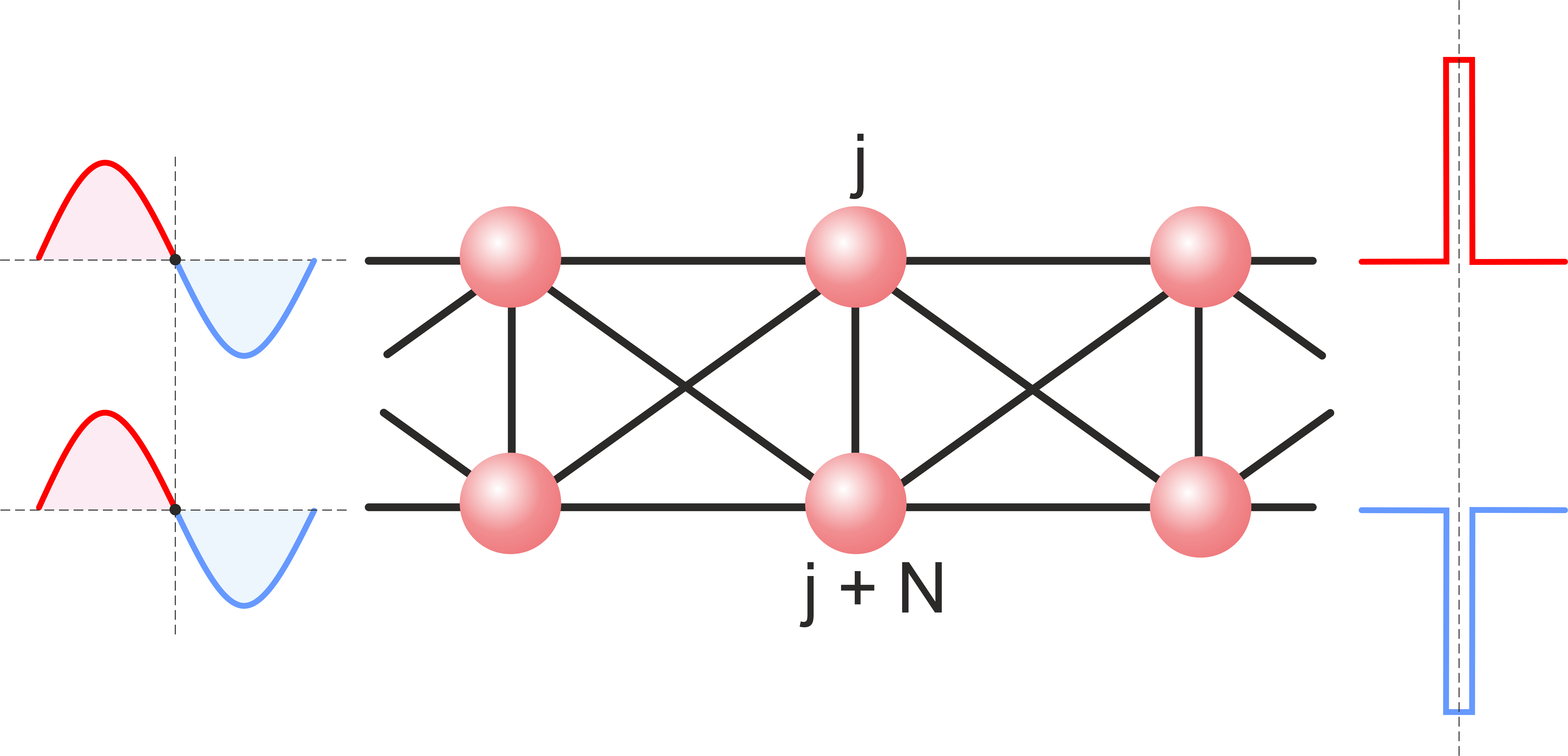}
(b) \includegraphics[width=0.9\columnwidth]{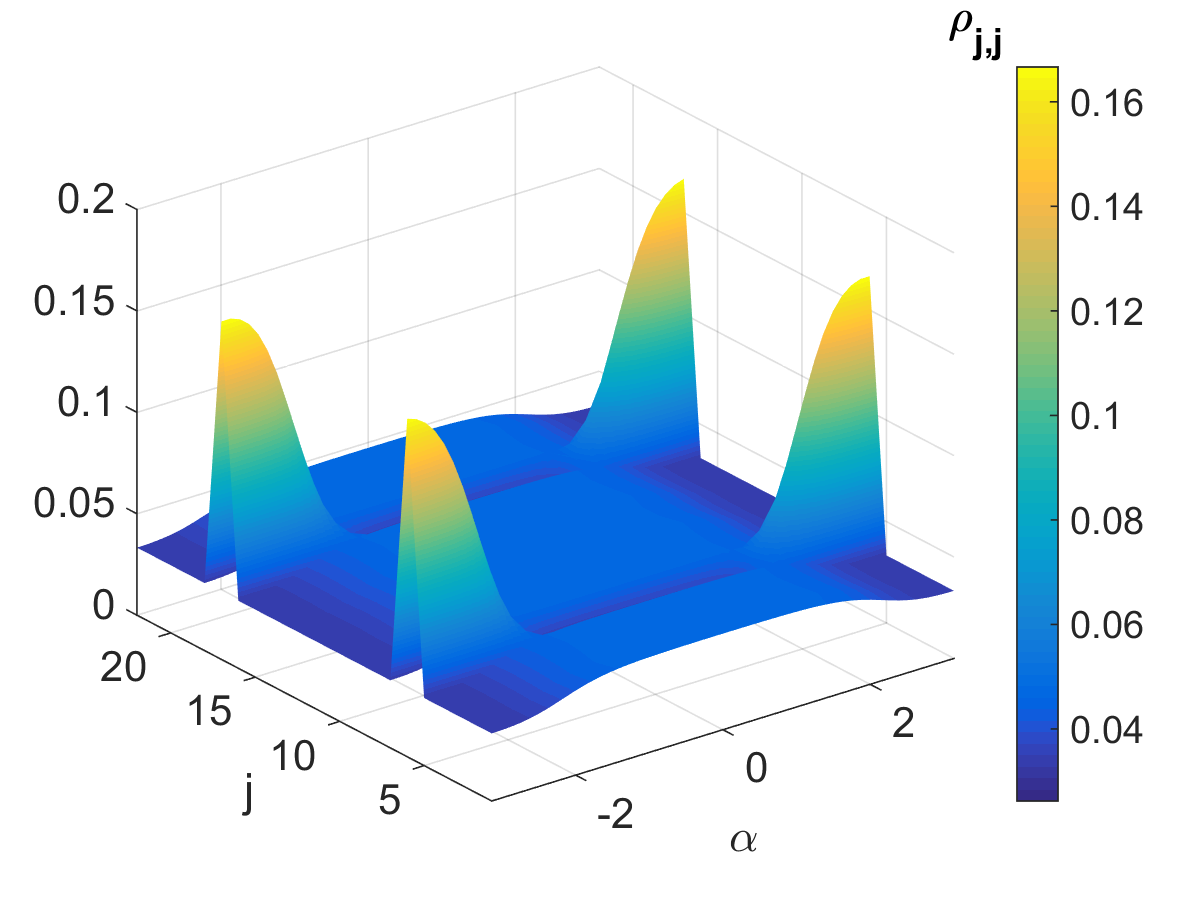}
\caption{(Color online) (a) Cross-stitch topology 
and schematic in-phase plain-wave solutions from the dispersive band (to the left) against anti-phase compact flat band modes (to the right). 
(b) Control of localization on flat bands. Asymptotic values of the diagonal elements of the density matrix as functions of the phase parameter $\alpha$. 
The parameters are  $N=11$, $\gamma_{d}=\gamma_{nn}=0.1$.}    
 \label{fig:1}
\end{figure}

\begin{figure}[t]
\includegraphics[width=0.9\columnwidth]{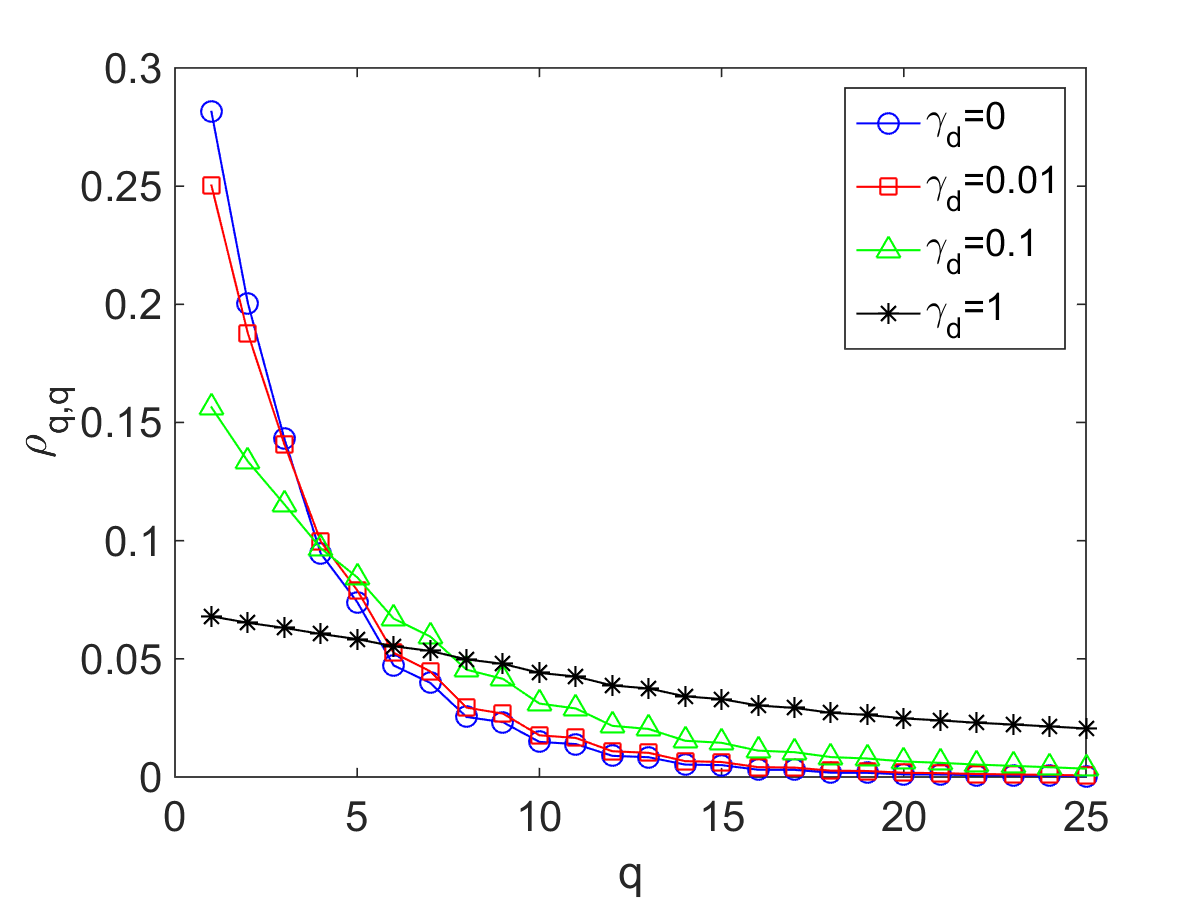}
\caption{(Color online) Anderson system: Diagonal elements 
of the asymptotic density matrix expressed in the eigenbasis of the Anderson Hamiltonian. The parameters are 
$\alpha=0, W=2, \gamma_{nn}=0.1, N=25$. The results are averaged over $N_r=10^3$ disorder realizations.}    
 \label{fig:1a}
\end{figure}

\begin{figure}[t]
\includegraphics[width=0.9\columnwidth]{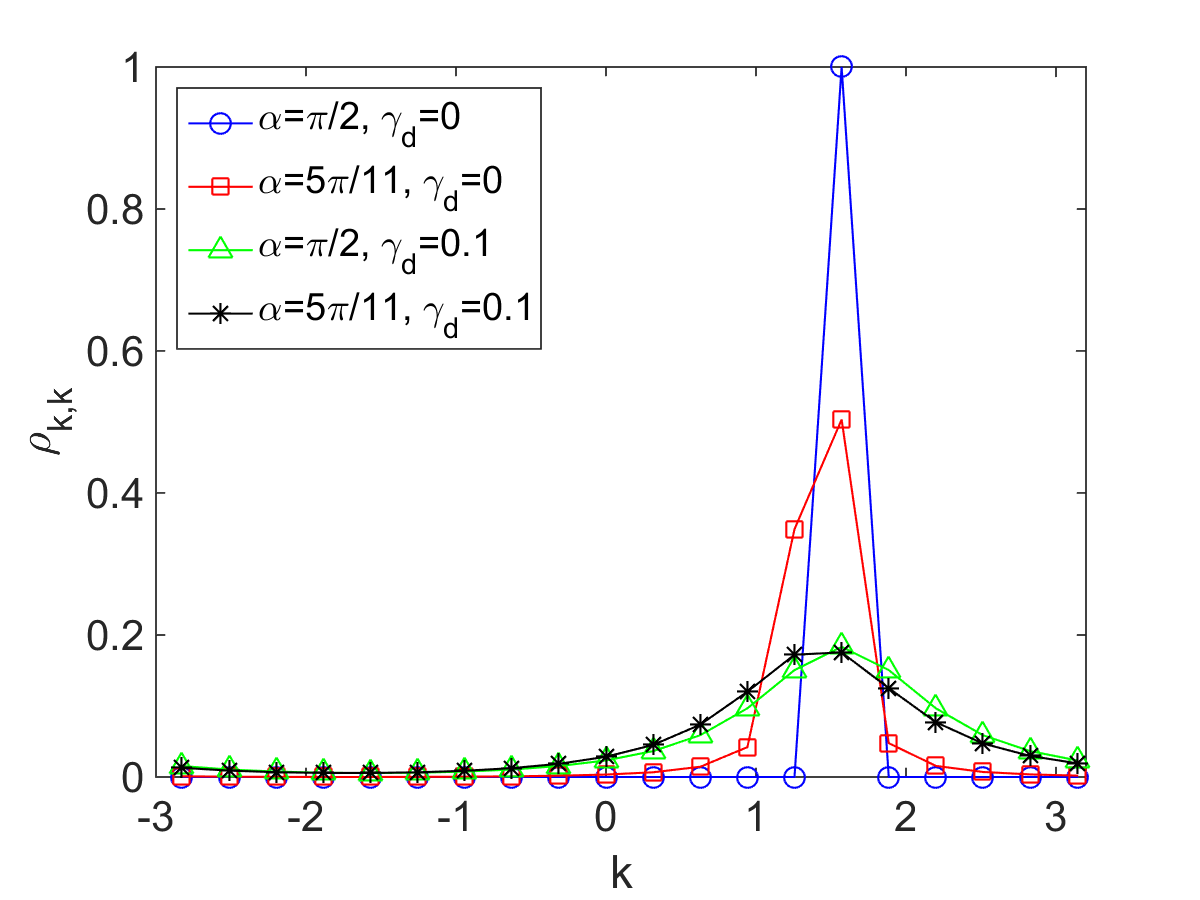}
\caption{(Color online) Disorder free system: Diagonal elements 
of the asymptotic density matrix expressed in the plain wave basis. The parameters are  $\gamma_{nn}=0.1$, and $N=20$. }    
 \label{fig:1b}
\end{figure}



We adress the asymptotic  solution $\varrho_{\infty}$ of Eq.~(\ref{eq:1}) only.
We assume that  due to the absence of relevant symmetries \cite{alicki,symmetry}, 
the  density operator $\varrho(t)$ is relaxing, under the action of propagator $\mathcal{P}_t$, towards a unique  operator (asymptotic state)
$\varrho_{\infty} = \lim_{t \rightarrow \infty} \mathcal{P}_t \varrho_0$ for all $\varrho_0$. Another words, it is
the unique kernel of the  Lindblad generator, $\mathcal{L}(\varrho_{\infty}) = 0$.
To find it $\varrho_{\infty}$ in a specified basis,  
we use a column-wise vectorization of the  density matrix and obtain the asymptotic solution, a $N^2$ vector $\rho_{\infty}$,  as the kernel of the 
Liouvillian-induced $N^2 \times N^2$ matrix $\Pi$,  $\Pi\rho_{\infty} = 0$. 
After folding the obtained vector back into the matrix form  and trace-normalizing it, 
we end up with the asymptotic state density matrix $\varrho_{\infty}$ \cite{exact}. 
We implenment both types of dissipative operators, Eqs.~(4-5),  simultaneously unless explicitly stated otherwise. 
Additionaly, in the case of Anderson Hamiltonian, we perform the averaging  over $N_r=10^3$ disorder realizations.

{\it Localization on a flat band. --} We first explore a possibility to 
shape a localized asymptotic states in a
flat-band lattice, where connection between the phase properties of the dissipator and 
resulting solution is more apparent. 
Specifically, we choose the ``cross-stitch'' flat-band topology \cite{Bergman2008,Richter2006}, 
that consists of two locally cross coupled chains of size $N$, see Fig.\ref{fig:1}(a). 
The disorder-free Hamiltonian, $\epsilon_j=0$, possesses a horizontal (flat) band, $E(k)=1$. 
The eigenstates on this band are localized at reciprocal sites $(j_0,j_0+N)$ from the upper and lower chains, 
with the wave function in anti-phase and zero elsewhere, $\psi_{j}=-\psi_{j+N}=\delta_{jj_0}/\sqrt{2}$, see sketch on the right part of Fig.\ref{fig:1}(a). 
The dispersive band, $E(k)=-1-4\cos(k), \ k=2\pi q/N, q=-N/2\ldots N/2$, holds plain waves $\psi_{j}=\psi_{j+N}=e^{i k j}/\sqrt{2N}$ 
which are in-phase in both chains, see sketch on the left  part of Fig.\ref{fig:1}(a).   

The control can be realized, for example, by using a single 
non-Hermitian local dissipator, Eq.(\ref{eq:4}), 
between the reciprocal sites $j=N/2$ and $n(j)=N/2+N$.
We expect, that the compact mode localized on the site $N/2$, $\psi_{j}=-\psi_{j+N}=\delta_{jN/2}/\sqrt{2}$, 
becomes a dark state of the dissipator for $\alpha=\pi$; at the same time, the in-phase modes on 
the dispersive band are subjected to dissipation. 
In turn, the whole dispersive band will become a dark state for $\alpha=0$, while the anti-phase compact mode at $j=N/2$ will near die out.  
 
 
To ensure robustness of dissipation-induced localization, 
we additionally introduce dephasing, by allowing  dissipators $V^{d}_j$, 
Eq. (\ref{eq:4a}), operate on every  site. Recall that in absence of $V^{nn}$, 
the asymptotic density operator would be the normalized identity. We set equal rates for both 
types of dissipators, $\gamma_{nn}=\gamma_d=0.1$.
\begin{figure*}[ht]
(a) \includegraphics[width=0.9\columnwidth]{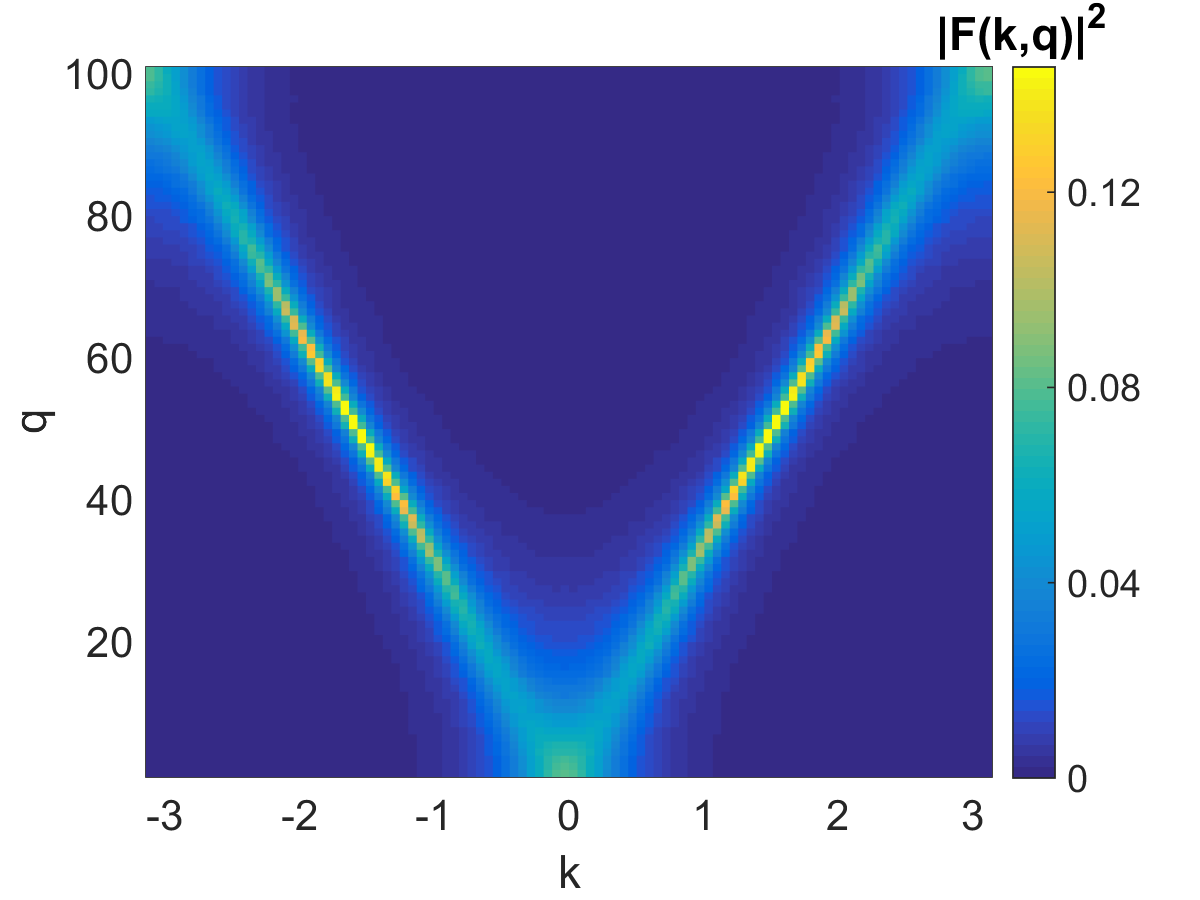}
(b) \includegraphics[width=0.9\columnwidth]{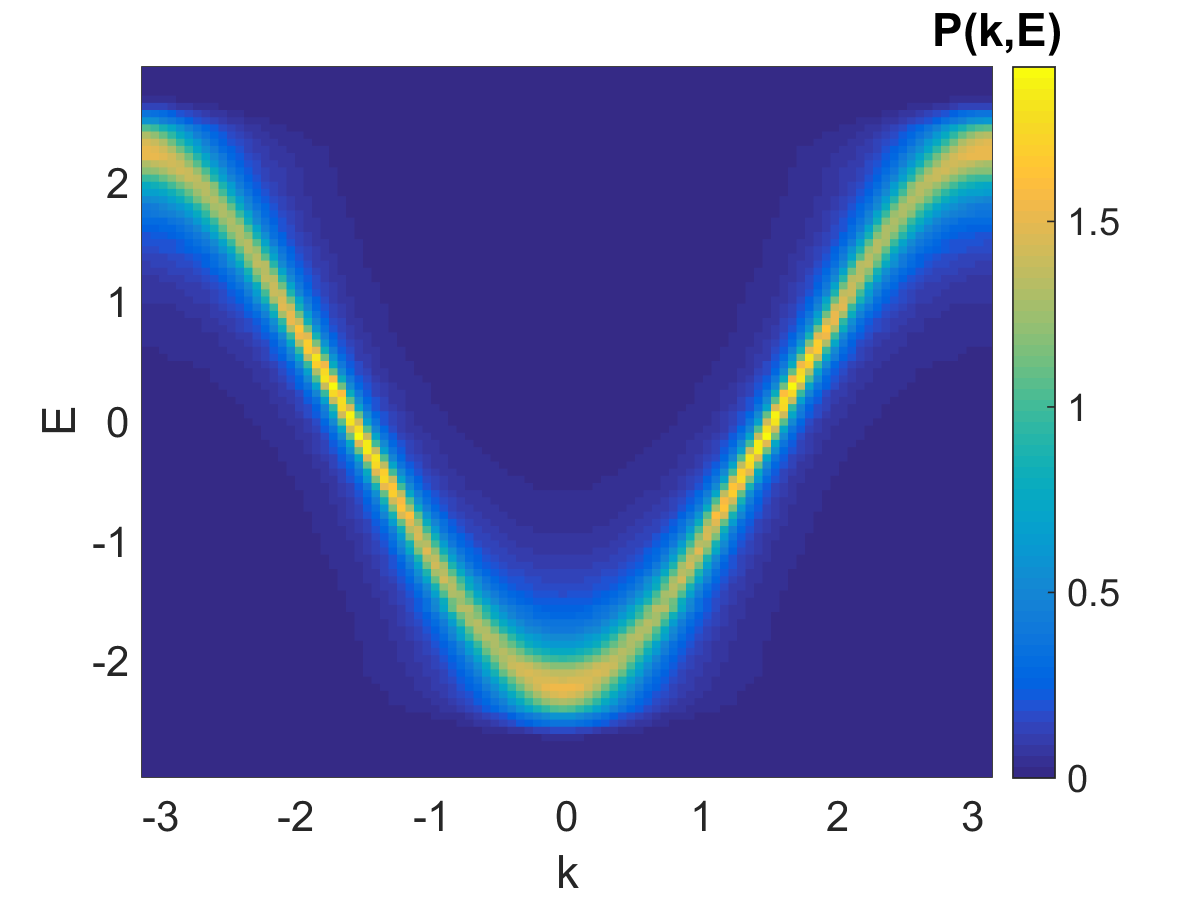}\\
(c) \includegraphics[width=0.9\columnwidth]{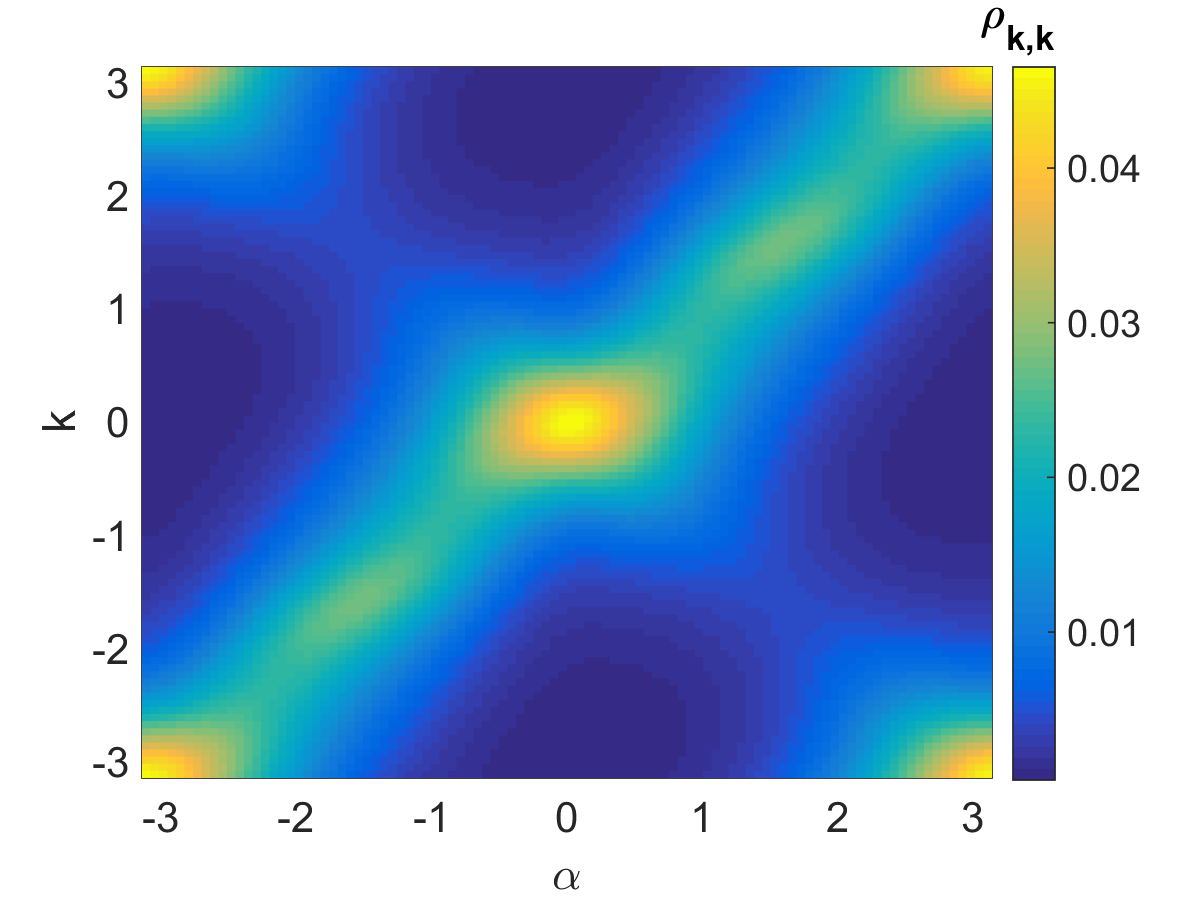}
(d) \includegraphics[width=0.9\columnwidth]{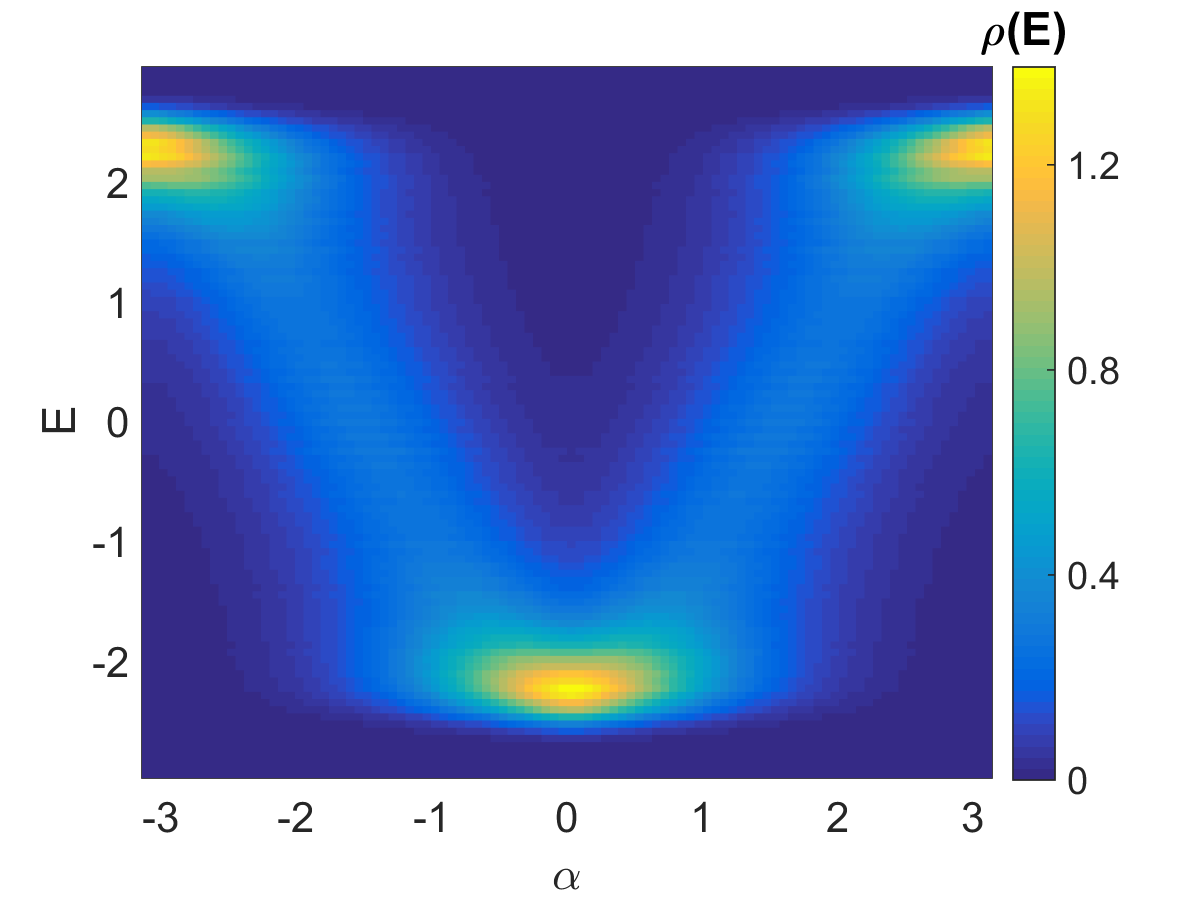}\\
\caption{(Color online) Anderson modes in Fourier space: Color coded (a) spatial harmonics $|F(k,q)|^2$ for wave number $k$ vs. mode number $q$ and (b) 
their spectral density $P(k,E)$. Asymptotic state of the open Anderson system: Color coded diagonal elements (averaged over $N_r=10^3$ disorder realizations)
of an asymptotic density matrix $\rho_{q,q}$ in the Fourier (c) and 
Anderson (d) basis controlled by the phase of dissipators $\alpha$. The parameters are $W=2, \gamma=0.1, N=100$.}    
\label{fig:2}
\end{figure*}
Now we consider  the dependence of the asymptotic state on the value of the parameter $\alpha$, Fig.\ref{fig:1}(b). 
As expected, we observe domination of the compact flat band mode in the central pair of sites for anti-phase local dissipation. 
However, the effect is pronounced in the much broader range, $|\alpha|\in(\pi/2,\pi]$. Remarkably, although the other flat band modes are not affected by this dissipative operator (formally they are  its dark states for any value of $\alpha$), 
they are also effectively suppressed. 
This can be understood as domination of the central compact mode due to the dissipation-induced `pumping' from dispersive modes. 
In the other parameter region, $|\alpha|\in[0,\pi/2)$, the asymptotic state looks like an almost homogeneous distribution 
with a noticeable drop corresponding to the suppressed central flat-band mode.   

The approach permits a straightforward extension by subjecting several sites to local non-Hermitian dissipation, 
promoting or suppressing respective compact modes. Ultimately, one has a possibility to arrange `dark bands', the 
mutually exclusive excitation of the whole flat and dispersive bands, by setting pairwise dissipators at each vertical pair, 
and choose $\alpha=0$ (dispersive regime) and $\alpha=\pi$ (zero dispersion). The result is exact in absence of Hermitian dissipators, $\gamma_d=0$, and approximative otherwise.   

{\it Anderson localization in the presence of dissipation. --} Next we consider the disorder-induced localization 
in the lattice described by Eq.~(\ref{eq:1}) with the 
Hamiltonian \cite{anderson} 
\begin{align}
H=&\sum_j \epsilon_j b_j^{\dagger}b_j -(b_{j}^{\dagger} b_{j+1} + b_{j+1}^{\dagger} b_{j}), 
\label{eq:5}
\end{align}
where $\epsilon_k\in\left[-W/2, W/2\right]$ are random uncorrelated on-site energies, and $W$ is the 
disorder strength.
We recall that the eigenvalues of the Hamiltonian are restricted to a finite interval, $E_q \in \left[-2-W/2, 2+W/2 \right]$, while the 
respective eigenstates, $A_j^{(q)}$,  are exponentially localized.
The localization length is approximated by $\xi_E \approx 24(4-E^2)/W^2$  \cite{Thouless1979}, with some corrections near the band edges \cite{Derrida}.

\begin{figure*}[ht]
(a) \includegraphics[width=0.9\columnwidth]{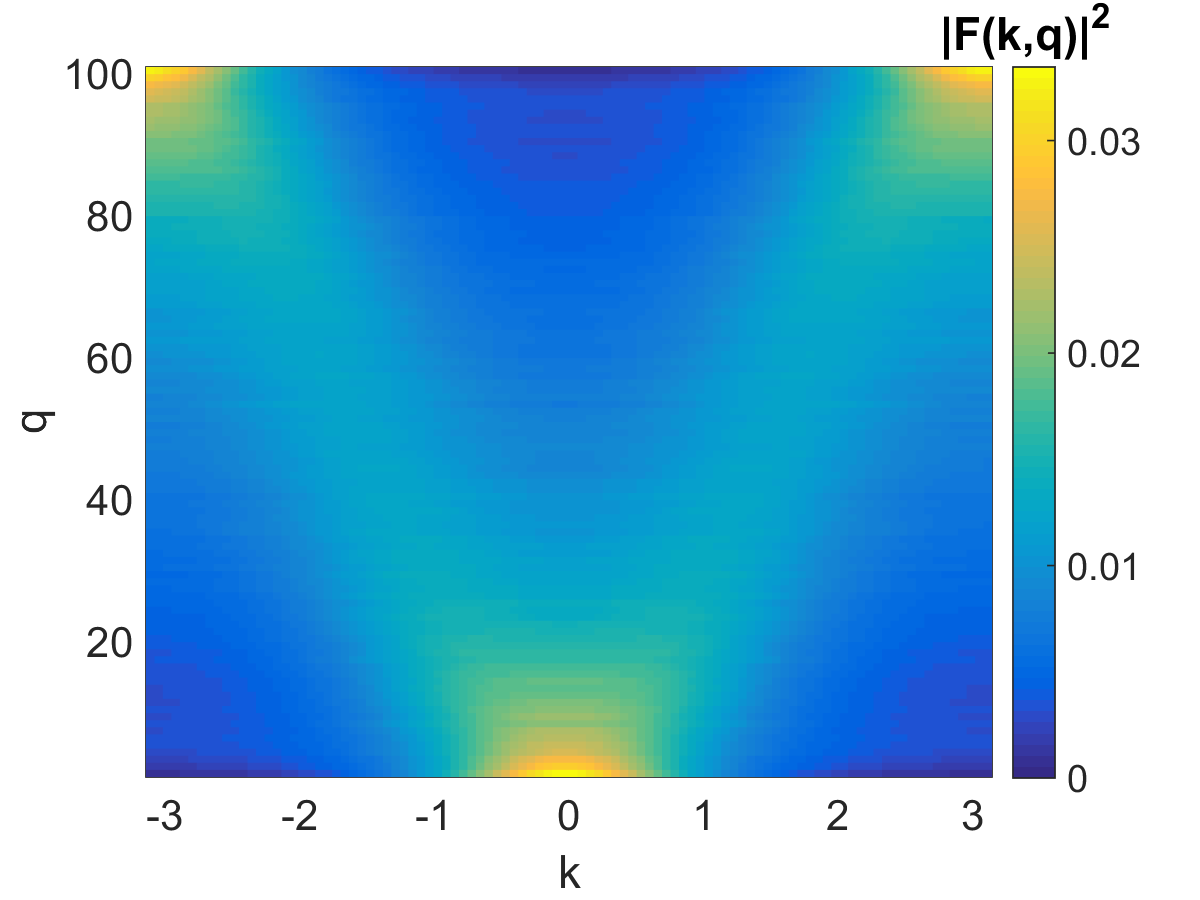}
(b) \includegraphics[width=0.9\columnwidth]{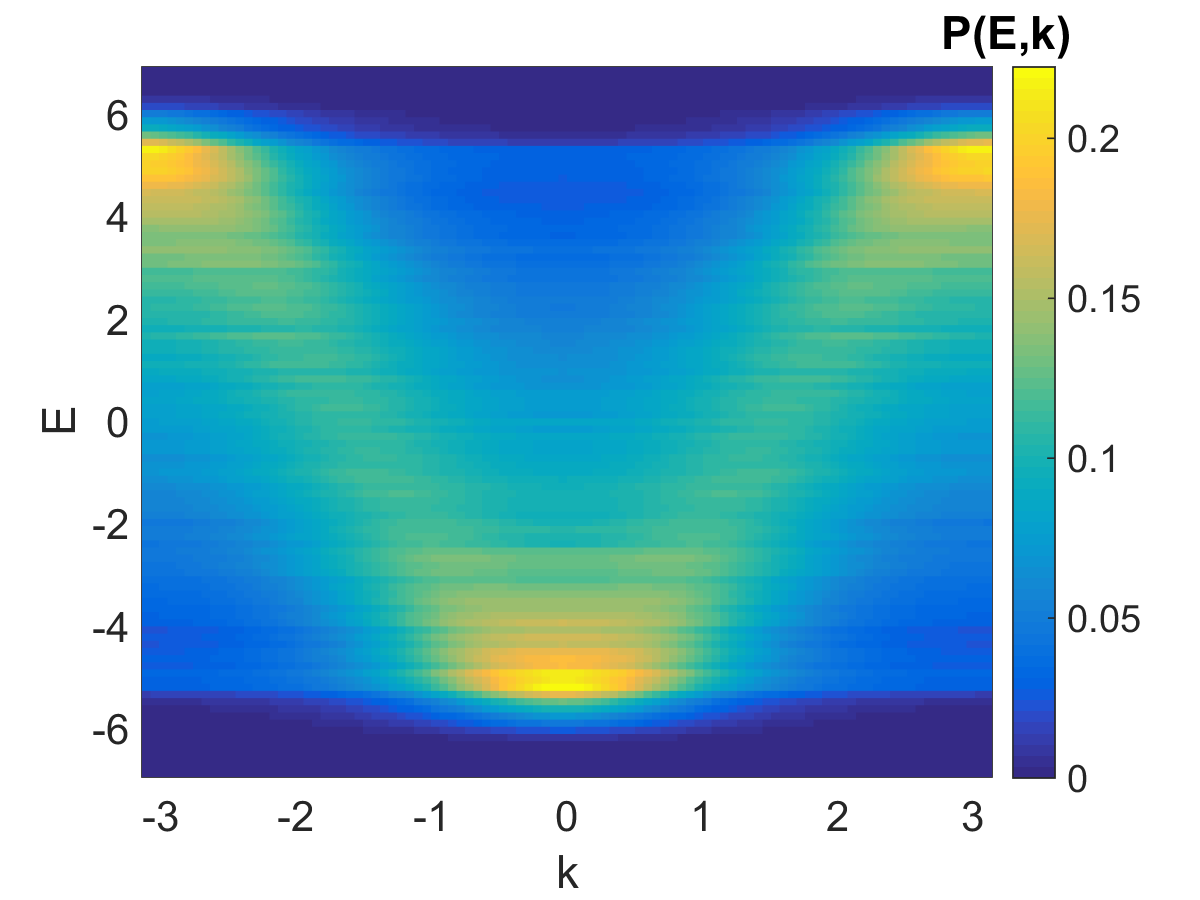} \\
(c) \includegraphics[width=0.9\columnwidth]{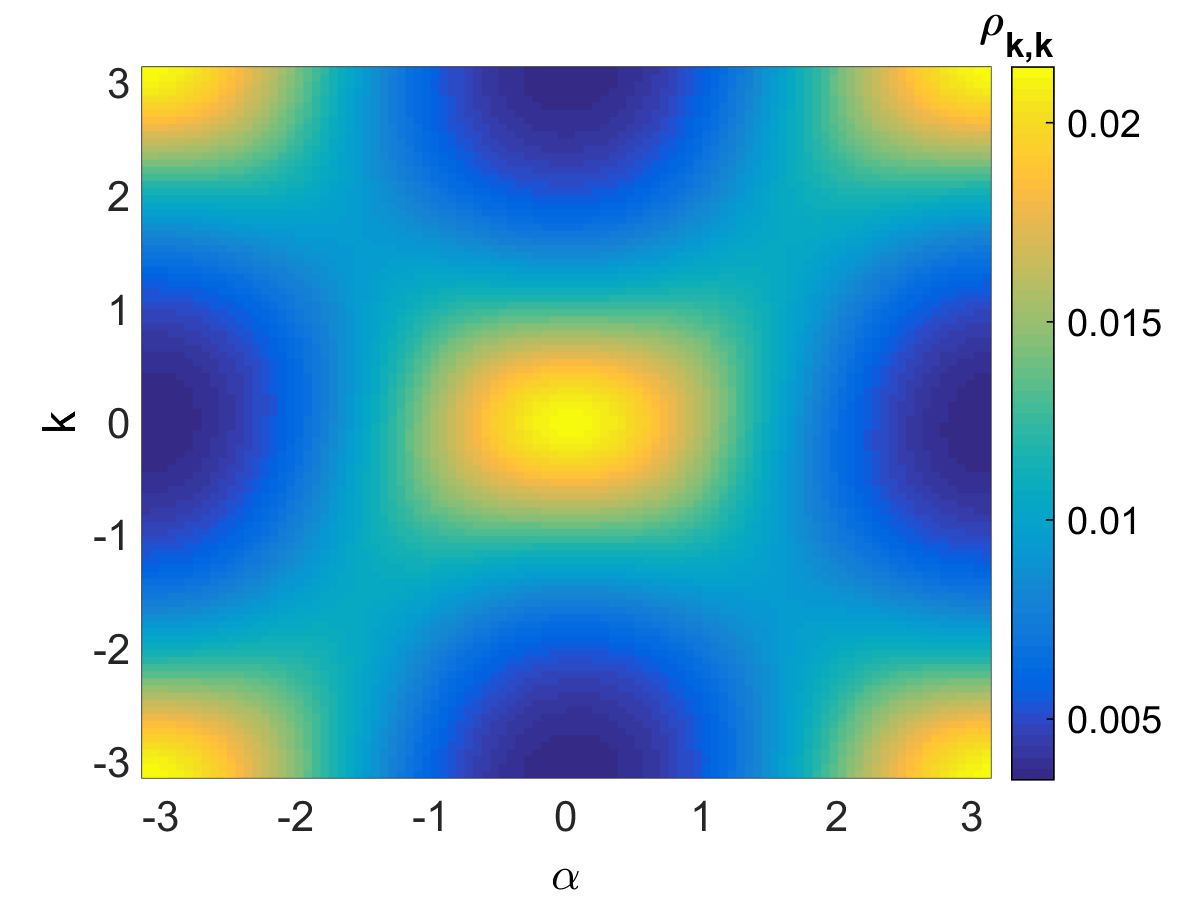}
(d) \includegraphics[width=0.9\columnwidth]{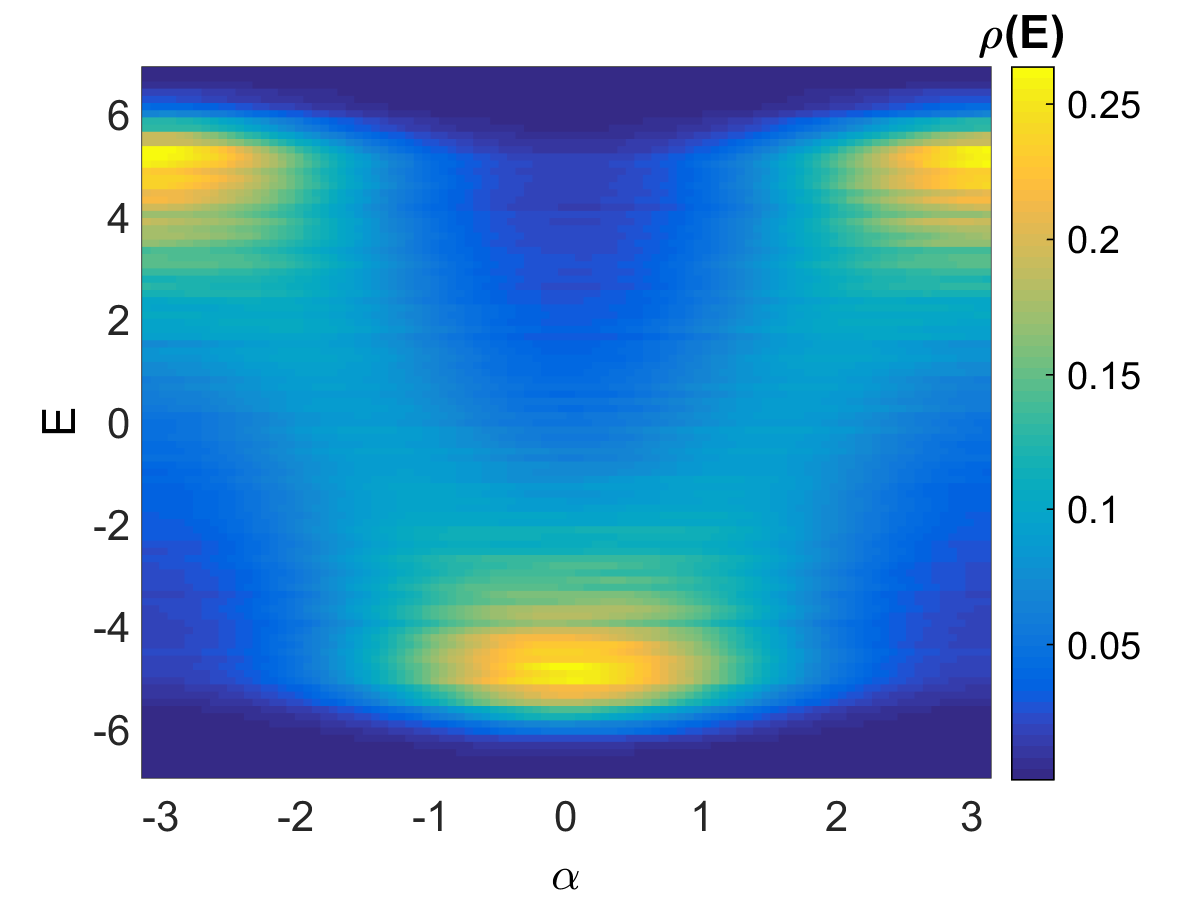}\\
\caption{(Color online) Anderson modes in Fourier space: color coded (a) 
spatial harmonics $|F(k,q)|^2$ for wave number $k$ vs. mode number $q$ and (b) their spectral density $P(E,k)$. 
Asymptotic state of the open Anderson system: Color coded diagonal elements (averaged over $N_r=10^3$ disorder realizations)of an asymptotic density matrix $\rho_{q,q}$ in the Fourier (c) and Anderson (d) basis controlled 
by the phase of dissipators $\alpha$. The parameters are $W=10, \gamma_{nH}=\gamma_H=0.1, N=100$.}    
\label{fig:3}
\end{figure*}

In Ref.~\cite{Yusipov2017}, it was demonstrated that pairwise dissipators, Eq. (5),  can favor specific Anderson modes. 
That is, the asymptotic state of the system is composed of Anderson modes, selected from around of a specific point in the spectrum. 
For example, the in-phase dissipation, $\alpha=0$, favors localization near the lower band edge, see Fig.\ref{fig:1a}). 
For $\gamma_d=0$, matrix elements of the corresponding  asymptotic density operator expressed  in the Anderson basis, $\varrho_{q,q}$, can be approximated as \cite{Yusipov2017}
\begin{equation}
\label{eq:6}
{\varrho}_{q,q}\propto\frac{2-E_q}{2+E_q}.
\end{equation}
Instead, the anti-phase dissipation, $\alpha=\pi$, facilitate localization near the upper band edge. 
Moreover, by using the next-to-nearest interaction, $n(j)=j+2$, it is possible to obtain an asymptotic state that consists of the Anderson modes from the spectrum center. 
In contrast to the two previous cases, the asymptotic density matrix in the direct basis does not  manifest (visually) localization, 
as the more extended modes from the middle of the spectrum substantially overlap in the space \cite{Yusipov2017}.

In Ref.~\cite{Yusipov2017} we have also conjectured that the preference by  
pairwise dissipators for specific  Anderson modes is due to the 
spatial-phase properties of the latter. More precisely, it is the phase structure which modes inherit from the 
seeding plane waves, the eigenstates of the Hamiltonian in the zero disorder limit \cite{ishii}. 
Here we corroborate the conjecture.

But first we demonstrate the  robustness of the phenomenon against local dephasing. 
We use $\alpha=0$ as a reference case, set the rate $\gamma_{nn}=0.1$ for the pairwise dissipators, and vary the rate of dephasing, $\gamma_d$. 
It turns out that not only weak dephasing, $\gamma_d\ll\gamma_{nn}$, does not noticeably change the structure of an asymptotic state but even
strong dephasing, $\gamma_d\gg\gamma_{nn}$, is not able to destroy the spectral  localization, see Fig.\ref{fig:1a}. 

Next we consider the zero-disorder limit in more details. 
There the basis of the Hamiltonian is formed by the plain waves $\psi_{j}=e^{i k j}/\sqrt{N}$, with the spectrum $E(k)=-2\cos(k), \ k=2\pi q/N, q=-N/2\ldots N/2$. 
It is straightforward to see that for specific values of the phase parameter, $\alpha=2\pi q/N$, 
the plain wave with respective $k=\alpha$ becomes a dark state of all $V^{nn}_j$, while all other eigenstates do not. 
In this case (and under assumption of zero dephasing, $\gamma_d=0$), the plain wave with $k=\alpha$ is the asymptotic state of 
the open system. 
We find that when $\alpha$ does not coincide with one of the wave vectors or dephasing  is present, the asymptotic state remains very close to the former dark state, 
with the plain waves with $k\approx\alpha$ contributing most, see Fig.\ref{fig:1b}.        

Disorder leads to Anderson localization of all eigenstates, 
also taking them away from the pool of dark states of $V^{nn}_j$, Eq.(\ref{eq:4}). 
However, it is known that Anderson modes inherit  phase properties of the original plain waves (at least in the regime of  weak disorder), 
although their amplitudes decay exponentially in space \cite{ishii}. Therefore, selective effect of local dissipators persists. 

We analyze  the structure of Anderson modes, $A_k^{(q)}$, in the plain wave (Fourier) basis for different disorder strength. 
Note, that while exponential localization in the direct space assumes de-localization in the plane wave basis, 
it does not exclude inhomogeneity of the distribution in there. 
The expansion coefficients, $F(k,q)=\sum A_k^{(q)}e^{i k j}/\sqrt{N}$, 
have pronounced maxima along the linear dependence, $q\propto\pm k_{max}$, see Fig.\ref{fig:2}(a). 
We also calculate the spectral density of expansion coefficients, 
\begin{equation}
P(k,E)=\lim\limits_{\Delta E\rightarrow 0}\frac{1}{\Delta E}\sum\limits_{q:E(q)\in[E,E+\Delta E]}|F(k,q)|^2,
\label{eq:7}
\end{equation}  
which closely reproduces dispersion relation for disorder-free system, Fig.\ref{fig:2}(b). 
It is noteworthy that these features are present  even in the strong disorder regime, $W=10$, see Fig.\ref{fig:3}(a,b).

The relation between the spatial structure of Anderson modes and their position in the spectrum gives a clue about the way to select the modes to form the asymptotic state.
Evidently, this can be done by varying the phase parameter $\alpha$ of pairwise dissipators. Numerical results reveal a well-shaped maximum in the diagonal elements of the asymptotic density 
expressed in the plain wave basis, $\rho_{k,k}$, such that $k_{max}\propto\alpha$, see Fig.\ref{fig:2}(c). 
We also calculate the spectral density of the diagonal elements in the Anderson basis, $\rho_{q,q}$, 
\begin{equation}
\rho(E)=\lim\limits_{\Delta E\rightarrow 0}\frac{1}{\Delta E}\sum\limits_{q:E(q)\in[E,E+\Delta E]}\rho_{q,q}.
\label{eq:8}
\end{equation}  
Plotted as a function of $\alpha$, it reveals the region of the Anderson spectrum whose modes contribute most to  the asymptotic state, Fig.\ref{fig:2}(d). 
Similarly, through with less sharp results,  this control recipe can can be used in the case of strong disorder, $W=10$; see Fig.\ref{fig:3}(c,d).

{\it Conclusions. --} We demonstrated that synthetic dissipation can
be used to control localization properties of the asymptotic states of single-particle quantum systems. 
The control mechanism  relies on the phase properties of localized modes of the system Hamiltonian, 
so that the modes appear dark  (or near dark) states of synthetic dissipators. 
Our  findings are relevant to a broad range of single-particle systems; these are other classes of flat bands, disordered flat band lattices, 
quasiperiodic (Aubry-Andre) potentials (for both, localization in direct and momentum space). Finally, we would like to speculate about perspective of using  the idea 
in the context of  many-body localization \cite{fish,les,les2,lazz}. However, this perspective, though very intriguing,  is  ambiguous at the moment. 

{\it Acknowledgments. --} This work was supported by the Russian Science Foundation grant No.\ 15-12-20029. 


\begin{thebibliography}{0}
\expandafter\ifx\csname natexlab\endcsname\relax\def\natexlab#1{#1}\fi
\expandafter\ifx\csname bibnamefont\endcsname\relax
  \def\bibnamefont#1{#1}\fi
\expandafter\ifx\csname bibfnamefont\endcsname\relax
  \def\bibfnamefont#1{#1}\fi
\expandafter\ifx\csname citenamefont\endcsname\relax
  \def\citenamefont#1{#1}\fi
\expandafter\ifx\csname url\endcsname\relax
  \def\url#1{\texttt{#1}}\fi
\expandafter\ifx\csname urlprefix\endcsname\relax\def\urlprefix{URL }\fi
\providecommand{\bibinfo}[2]{#2}
\providecommand{\eprint}[2][]{\url{#2}}

\end{thebibliography}


\begin{thebibliography}{1000}




\bibitem{anderson} P.W. Anderson, Phys. Rev. \textbf{109}, 1492 (1958).

\bibitem{Kramer1993} B. Kramer and A. MacKinnon, Rep. Prog. Phys. {\bf 56}, 1469 (1993).
\bibitem{Evers2008} F. Evers and A. Mirlin, Rev. Mod. Phys. {\bf 80}, 1355 (2008).
\bibitem{fifty} 50 Years of Anderson Localization, ed. by E. Abrahams (World Scientific, 2010).


\bibitem{fifty2} A. Lagendijk, B. van Tiggelen, and D. S. Wiersma, Physics Today \textbf{62}, 24 (2009).


\bibitem{Segev2013}  M. Segev, Y. Silberberg, and D. N. Christodoulides, Nature Photon. \textbf{7}, 197 (2013).
\bibitem{Billy2008} J.~Billy, V.~Josse, Z.~Zuo, A.~Bernard, B.~Hambrecht, P.~Lugan, D.~Cl\'{e}ment, L.~Sanchez-Palencia, P.~Bouyer, and A.~Aspect, Nature {\bf 453}, 891 (2008).
\bibitem{Roati2008} G.~Roati, C.~D'Errico, L.~Fallani, M.~Fattori, C.~Fort, M.~Zaccanti, G.~Modugno, M.~Modugno, and M.~Inguscio, Nature {\bf 453}, 895 (2008).
\bibitem{Kondov2011} S. S.~Kondov, W. R.~McGehee, J. J.~Zirbel, B.~DeMarco, Science {\bf 334}, 66 (2011).
\bibitem{Jen2012} F.~Jendrzejewski,	A.~Bernard,	K.~M\"{u}ller,	P.~Cheinet,	V.~Josse,	M.~Piraud,	L.~Pezz\'{e},	L.~Sanchez-Palencia,	A.~Aspect, and P.~Bouyer, Nature Phys. {\bf 8}, 398 (2012).





\bibitem{book} H.-P. Breuer, F. Petruccione, \textit{The Theory of Open Quantum Systems} (Oxford University Press, Oxford, 2002).


\bibitem{Bergman2008} D.L. Bergman, C. Wu and L. Balents, Phys. Rev. B {\bf 78}, 125104 (2008).
\bibitem{Richter2006} O. Derzhko and J. Richter, Eur. Phys. J. B {\bf 52}, 23 (2006).
\bibitem{Flach2014} S. Flach, D. Leykam, J.D. Bodyfelt, P. Matthies, and A.S.
Desyatnikov, Europhys. Lett. {\bf 105}, 30001(2014); {\bf 106}, 19901 (2014).






\bibitem{DiehlZoller2008} S. Diehl, A. Micheli, A. Kantian, B. Kraus, H. P. B\"uchler, P. Zoller, Nature Physics {\bf 4}, 878 (2008).
\bibitem{KrausZoller} B. Kraus H. P. B\"uchler, S. Diehl, A. Kantian, A. Micheli, P. Zoller, Phys. Rev. A {\bf 78}, 042307 (2008).

\bibitem{wolf2009} F.~Verstraete, M. M.~Wolf, and J.I.~Cirac, Nature Phys. {\bf 5}, 633 (2009).

\bibitem{Stano2013} P.~Stano and P.~Jacquod, Nature Photonics \textbf{7}, 66 (2013).
\bibitem{LiuJ.2014} J.~Liu,	P. D.~Garcia,	S.~Ek,	N.~Gregersen,	T.~Suhr,	M.~Schubert,	J.~M\o rk,	S.~Stobbe, and P.~Lodahl, Nat. Nanotech. \textbf{9}, 285 (2014).
\bibitem{ivanchenko2015a} T. V.~Laptyeva, A. A.~Tikhomirov, O. I.~Kanakov and M. V.~Ivanchenko, Sci. Rep. {\bf 5}, 13263 (2015).
\bibitem{ivanchenko2015b} T. V. Laptyeva, S. V. Denisov, G. V. Osipov, M. V.  JETP Lett., {\bf 102} (9), 603 (2015).

\bibitem{les0} S. Genway, I. Lesanovsky, and J. P. Garrahan, Phys. Rev. E \textbf{89}, 042129 (2014).
\bibitem{Yusipov2017} I. Yusipov, T. Laptyeva, S. Denisov, and M. Ivanchenko, Phys. Rev. Lett. {\bf 118}, 070402 (2017).
















\bibitem{alicki} R. Alicki, K. Lendi, 1987, \textit{Quantum Dynamical Semigroups and Applications},
Lecture Notes in Physics, Vol. 286 (Springer, Berlin).

\bibitem{fish} M. F. Fisher, M. Maksymenko, E. Altman, Phys. Rev. Lett. {\bf 116}, 160401 (2016).
\bibitem{les} E. Levi, M. Heyl, I. Lesanovsky, J. P. Garrahan, Phys. Rev. Lett. {\bf 116}, 237203 (2015).
\bibitem{les2} B. Everest, I. Lesanovsky, J.P. Garrahan, E. Levi, Phys. Rev. B {\bf 95}, 024310 (2017).
\bibitem{lazz} A. Lazarides and R.  Moessner, Phys. Rev. B \textbf{95}, 195135 (2017).

\bibitem{marcos2012} D. Marcos, A. Tomadin, S. Diehl, and P. Rabl, New J. Phys. \textbf{15}, 055005 (2012).



\bibitem{QO} P. Meystre and M. Sargent, \textit{Elements of Quantum Optics} (Springer, Berlin, 4 ed., 2007).
 

\bibitem{symmetry}  V. V.~Albert, L.~Jiang, Phys. Rev. A {\bf 89}, 022118 (2014).

\bibitem{exact}  Maximal absolute value of the elements in the r.h.s of equation (\ref{eq:1}) after substituion 
of $\varrho_{\infty}$ does not exceed $10^{-14}$.


\bibitem{Thouless1979} D. J.~Thouless, In: \textit{Ill-condensed Matter}, Eds. R. Balian, R. Maynard, and G. Toulouse
(North-Holland, 1979).
\bibitem{Derrida} B. Derrida and E. Gardner, J. Physique {\bf 45}, 1283 (1984).
              
\bibitem{ishii} K. Ishii, Prog. Theor. Phys. Suppl. \textbf{53}, 77 (1973).












































 
























\end{thebibliography}
\end{document}